\documentclass[aps,prl,reprint,preprintnumbers]{revtex4-1}
\usepackage{graphicx}
\usepackage{bm}
\usepackage{epsfig}
\usepackage{ulem}
\usepackage{color}
\usepackage[table]{xcolor}
\usepackage{colortbl}
\usepackage{mathrsfs}
\usepackage{dcolumn}
\usepackage{setspace}
\usepackage{array}
\usepackage{amsmath}
\usepackage{amssymb}
\usepackage{dsfont}
\usepackage{gensymb}
\usepackage{dcolumn}
\usepackage{multirow}
\usepackage{bibentry,natbib}
\usepackage{booktabs}
\usepackage{hyperref}
\hypersetup{colorlinks=true, linkcolor=blue, filecolor=blue,  urlcolor=blue}
\definecolor{lightgray}{gray}{0.9}
\begin{document}
\bibliographystyle{unsrt}

\title{Axial Casimir Force}

\author{Qing-Dong Jiang$^1$, Frank Wilczek${^1}{^2}{^3}{^4}$}
\affiliation{{}\\ $^1$Department of Physics, Stockholm University, Stockholm SE-106 91 Sweden\\
$^2$Center for Theoretical Physics, Massachusetts Institute of Technology, Cambridge, Massachusetts 02139 USA\\
$^3$Wilczek Quantum Center, Department of Physics and Astronomy, Shanghai Jiao Tong University, Shanghai 200240, China\\
$^4$Department of Physics and Origins Project, Arizona State University, Tempe AZ 25287 USA}

\begin{abstract}
Quantum fluctuations in vacuum can exert a dissipative force on moving objects, which is known as Casimir friction. Especially, a rotating particle in the vacuum will eventually slow down due to the dissipative Casimir friction.  Here, we identify a \textit{dissipationless} force by examining a rotating particle near a bi-isotropic media that generally breaks parity symmetry or/and time-reversal symmetry. The direction of the dissipationless vacuum force is always parallel with the rotating axis of the particle.   We therefore call this dissipationless vacuum force the axial Casimir force.
\end{abstract}
\preprint{MIT-CTP/6063}
\maketitle

\section{I. Introduction}
Originating from quantum fluctuations, the Casimir effect describes the phenomenon where an attractive force emerges between two non-contacted, uncharged plates in vacuum \cite{Casimir}. The Casimir effect tells us that vacuum is not empty, but full of fluctuations with photons popping in and out. In fact, there are many other effects that can manifest the fluctuating nature of vacuum. For example, quantum fluctuations can exert a torque on bodies that lack rotational symmetry, called Casimir torque \cite{YBarash,measureCasimirtorque}. If some discrete symmetries are broken in materials, quantum fluctuation can transmit symmetry breaking effect to nearby atoms and perturbs the atom's spectra, namely the quantum atmosphere effect \cite{jiangWilczek}.  
In recent years, another interesting phenomenon, called Casimir friction, was discovered. Here, objects moving relative to each other can feel a dissipative viscous force due to the exchange of Doppler-shifted photons \cite{JBPendry}.  Perhaps counter-intuitively, a spinning object in vacuum will eventually slow down due to Casimir friction \cite{AManjavacas1}. In recent years, theorists have proposed many models that feature the Casimir friction \cite{AIVolokitin}, and some of them are closely related to experimental phenomena \cite{MOigrebinskii,AVolokitin2}. 

However, to our best knowledge, all the proposed Casimir friction phenomena (motion-induced vacuum forces) are dissipative. A natural question then arises: is it possible to find a dissipationless motion-induced vacuum force? This question is partially motivated by the recent progress in quantum Hall physics, where dissipationless Hall viscosity emerges as a new topological signature \cite{JAvron}.  We address this question in this paper by examining a rotating particle near a bi-isotropic material  (BIM) plate.  Existing commonly in nature, BIMs include materials that break time-reversal symmetry (TRS) or parity symmetry (PS) or both (PTS) \cite{ILindell}. In recent years, the widely studied Chern insulators \cite{FHaldane} and chiral metamaterials \cite{STretyakov} can be classified as bi-isotropic materials breaking TRS and PS, respectively.  

We show that, in addition to the dissipative Casimir friction, a dissipationless force can emerge for a rotating particle near a PS or TS (or both) breaking BIMs.  Since the dissipationless  rotation-induced force is always parallel to the particle's rotation axis and changes sign when its spinning direction is reversed, we, therefore, call it the axial Casimir force (ACF). Two cases are of particular interest: (i) when the rotation axis is parallel to the BIM plate, the axial Casimir force is lateral (L-ACF); (ii) when the rotation axis is perpendicular to the BIM plate, the axial Casimir force is vertical (V-ACF) [Fig. 1]. We calculate ACF both numerically and analytically, and show that TS breaking is crucial for V-ACF, whereas, by contrast, PS breaking is important for L-ACF.  Let us observe that very recent experiments have already achieved a superfast rotation of nanoparticles, making the ACF within the experimental reach \cite{RReimann}.

\begin{figure}[!htb]
\centering
\includegraphics[height=3.7cm, width=7cm, angle=0]{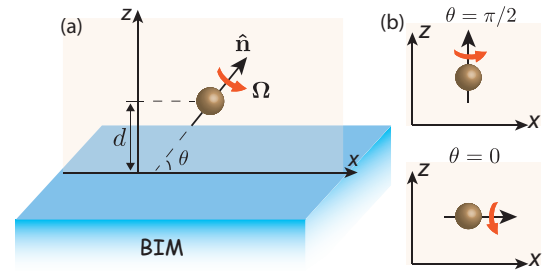}
\caption{Schematic of the structure.
$d$ is the distance from the center of the rotating object to the BIM plane. $\Omega$ represents the rotating frequency of the object. $\bold{\hat n}$ is the unit vector in the rotating direction.  $\theta$ is the angle between $\bold{\hat n}$ and the $x$ direction. (a) shows the general case, while (b) shows two special rotating directions, perpendicular to the BIM plane (top) and parallel with the BIM plane (bottom). \label{fig:1}}
\end{figure}

\section{II. Model}
We consider a spherical, isotropic particle rotating with frequency $\Omega$ located at the position $\bold r_0=(0,0,d)$ above a BIM plate at $z=0$ plane [Figure 1 (a)].  Without loss of generality, we assume that the rotating axis lies in the $x$-$z$ plane and forms a $\theta$ angle with $x$ axis. In this paper, we study the dissipationless ACF along the rotating axis $\bold{\hat n}$. Particularly, when $\theta=0$, the ACF lies in the $x$-direction, becoming a L-ACF;  when $\theta=\pi/2$, the ACF is parallel with $z$-direction, leading to the V-ACF. 

We assume that the particle is small enough so that it can be safely described by polarization function $\alpha(\omega)$, for instance a small metallic ball. (In other words, the size of the particle $\rm R$ is much smaller that the distance $d$.)  The electromagnetic force that exerts on an electric dipole in the direction $\bold{\hat n}$ can be evaluated via the formula
$
F_{n}=p_i(t)\partial_{\hat n} E_i(\bold r_0, t)
$ ($i\in \{x,y,z\}$),
where $p_i(t)$ and $E_i(\bold r_0,t)$ are, respectively, the instantaneous electric dipole moment at time $t$ and the electric field at the particle. (Einstein summation rule is implied through out this paper.) Note that we have omitted any magnetic dipole contribution, which is much smaller than the electric dipole contribution \cite{AVolokitin2}. We will further elaborate this point later in the paper. 
Although the average electric dipole and electric field are zero in vacuum, quantum fluctuation can still induce a instantaneous dipole, therefore exert a force on the particle. (This is also the mechanism of Wan der Vaals force.)
There are two kinds of fluctuations that contribute to the ACF: (i) fluctuations of the dipole moment of the particle, and (ii) fluctuations of the field caused by the electromagnetic response of the BIM plate.
Therefore, the total ACF includes two terms, i.e.,
\begin{eqnarray}\label{eq1}
F_{n}=\langle p_i^{fl}(t)\partial_{\hat n}E_i^{ind}(\bold r_0,t)\rangle+\langle p_i^{ind}(t)\partial_{\hat n}E_i^{fl}(\bold r_0,t)\rangle,
\end{eqnarray}
where $\langle \cdots \rangle$ stands for the average over fluctuations in vacuum.  In this formula $p_i^{fl (ind)}$ and $E_i^{fl(ind)}$ are, respectively, the fluctuating (induced) electric dipole moment and electric field at the particle.  (Note that the cross terms $\langle p^{fl}\partial_{\hat n} E^{fl}\rangle$  and $\langle p^{ind}\partial_{\hat n} E^{ind}\rangle$ vanish in vacuum because dipole moment and electric field arise from different sources.) When the particle  is not rotating, the force in z-direction is the usual Casimir-Polder force. As soon as the particle rotates, ACF will emerge, have an additional component in z-direction. 

Applying Fourier transformation, one can write down the induced field (dipole moment) in terms of the fluctuation of the dipole moment (electric field) in $\omega-$space, yielding
$E_i^{ind}(\bold r,\omega)=G_{ij}(\bold r,\bold r_0,\omega)  p_j^{fl}(\omega)$ and 
$p_i^{ind}(\omega)=\alpha_{ij}(\omega)  E_j^{fl}(\bold r_0,\omega)$,
where $G_{ij}$ and $\alpha_{ij}$ represent Green's tensor and polarization tensor, respectively.  Substitute the above equations into Eqn. \eqref{eq1}, and one can obtain
\begin{eqnarray}\label{eq4}
F_{\hat n}=\int_{-\infty}^{\infty} \frac{d\omega d\omega^{\prime}}{4\pi^2}\, e^{-i(\omega+\omega^{\prime})t}\left\{\langle  p_i^{fl} p_j^{fl}\partial_{\hat n}G_{ij}(\bold r_0,\bold r_0,\omega^{\prime})\rangle\right.\nonumber\\+\left.\langle  \alpha_{ij}(\omega)\partial_{\hat n}E_i^{fl}(\bold r_0,\omega^{\prime})E_j^{fl}(\bold r_0, \omega)\rangle\right\},\qquad
\end{eqnarray}
where one should notice that the derivative only acts on the first component of Green's tensor, i.e., $\partial_{\hat n}G_{ij}(\bold r_0,\bold r_0,\omega)\equiv \partial_{\hat n} G_{ij}(\bold r,\bold r_0,\omega)|_{\bold r=\bold r_0}$. We emphasize that, in Eqn. \eqref{eq4}, $p_i$ and $\alpha_{ij}$ are the effective electric dipole moment and electric polarizability in the laboratory frame, respectively. However, the electric dipole and polarizability are defined in the rotating frame of the particle. Therefore, one needs to identify the transformation from electric dipole or polarizability ($\tilde p_i$ or $\tilde \alpha_{ij}$) in the rotating frame to those in the laboratory frame \cite{AManjavacas1,AManjavacas}:
$p_i(\omega)=\Lambda_{ij}^+\tilde p_j(\omega_+)+\Lambda_{ij}^0\tilde p_j(\omega)+\Lambda_{ij}^{-} \tilde p_j(\omega_-)$ and 
$\alpha_{ij}(\omega)=\Gamma_{ijkl}^{+}\tilde \alpha_{kl}(\omega_+)+\Gamma_{ijkl}^{0}\tilde \alpha_{kl}(\omega)+\Gamma_{ijkl}^{-}\tilde \alpha_{kl}(\omega_-)$,
where $\omega_{\pm}=\omega\pm\Omega$ is the Doppler-shifted frequency due to rotation. Here, $\Lambda^{0}$, $\Lambda^{\pm}$ and $\Gamma^{0}$, $\Gamma^{\pm}$ represent the transformation tensor for dipole moment and polarizability. (See Appendix A.)

By applying fluctuation-dissipation theorem (FDT) to Eqn. \eqref{eq4},  we obtain a compact expression of  \textbf{axial Casimir force}:
\begin{eqnarray}\label{eq8}
F_{n}(\Omega)=F_x(\Omega)\cos^2\theta+F_z(\Omega)\sin^2\theta.
\end{eqnarray}
In this formula, $\theta$ denotes the rotating direction of the particle [Fig. 1(a)], $F_{x/z}$ denotes the ACF in x/z direction with expressions:
\begin{eqnarray}\label{eq9}
\begin{aligned}
F_{x/z}(\Omega)&=\frac{\hbar}{\pi}\int_0^{\infty} d\omega\,   \mathrm{Im} \left\{\Sigma_{x/z} \right\}\times \\&\left[\mathrm{Im}\,\alpha (\omega_+)N(\omega_+)- \mathrm{Im}\,\alpha (\omega_-)N(\omega_-)\right].
\end{aligned}
\end{eqnarray}
Here,  the differential Green's functions $\Sigma_{x/z}$ are determined by the surface Green's tensor $G_{ij}$ of the BIM plate via $\Sigma_x=\partial_xG_{yz}-\partial_xG_{zy}$ and $ \Sigma_z=\partial_zG_{xy}-\partial_zG_{yx}$;    $N(\omega_{\pm})\equiv n(T_1,\omega_\pm)-n(T_2,\omega)$ is defined by the difference of Bose-Einstein distribution, where $T_1$ and $T_2$ are temperatures at the rotating particle and the BIM plate, respectively. Note that, in deriving the above formula, we have used the isotropic assumption of the electric polarizability of the particle, i.e., $\tilde \alpha_{ij}(\omega)=\tilde \alpha(\omega) \delta_{ij}$ ($i,j\in \left\{x, y, z\right\}$).  Eqns. 
\eqref{eq8} and \eqref{eq9} are the main results of this paper. We stress that the ACF is different from the usual Casimir-Polder force, because ACF exists only when the particle is rotating with a finite speed.  We shall compare ACF with the usual Casimir-Polder force later in this paper and in the appendix. 

\section{III. Criterion of ACF - TRS/PS breaking}
In this part, we demonstrate that the emergence of an ACF requires TRS/PS breaking of the underlying BIM plate.  A BIM plate 
can generally be described by the constitutive relations $\bold D=\epsilon \bold E+(\chi-i\kappa)\sqrt{\epsilon_0\mu_0}\bold H$ and $\bold B=\mu\bold H+(\chi+i\kappa)\sqrt{\epsilon_0\mu_0}\bold E$, where $\epsilon$ ($\epsilon_0$) and $\mu$ ($\mu_0$) are, respectively, the permittivity and permeability of the BIM plate (vacuum). The essence of BIMs is encoded in the magnetoelectric parameters $\chi$ and $\kappa$, which characterize the non-reciprocity and chirality of the system, respectively. BIM with $\chi\neq 0$ and $\kappa=0$ has been called Tellegen medium, where TRS is broken. By contrast, BIM with $\chi= 0$ and $\kappa\neq 0$ has been labeled Pasteur medium, where PS is violated. Materials with $\chi=0$ and $\kappa=0$ is usually called simple isotropic medium, whereas, by contrast, both $\chi\neq 0$ and $\kappa\neq 0$ represent more general BIMs. [See Table I] 
\begin{table}
\centering
\caption{Classification of BIMs  \cite{ILindell} with axial Casimir force.}
\begin{tabular}{cccc}
\hline\hline
non-reciprocity & chirality & classification & \color{blue}{axial Casimir force}\\
\hline
\rowcolor[gray]{0.88}
$\chi=0$ & $\kappa=0$ & simple BIM & $F_{n}=0$\\
\hline
\rowcolor[gray]{0.95}
$\chi=0$ & $\kappa\neq 0$ & Pasteur & $F_x \neq 0$; $F_z=0$\\
\hline
\rowcolor[gray]{0.88}
$\chi\neq 0$ & $\kappa=0$ & Tellegen & $F_x= 0$; $F_z\neq  0$\\
\hline
\rowcolor[gray]{0.95}
$\chi\neq 0$ & $\kappa\neq 0$ & general BIM & $F_{x}\neq 0$; $F_{z}\neq 0$\\
\hline
\end{tabular}
\end{table}

With the constitutive relations of BIMs, one can study the electromagnetic response of BIMs. To obtain the ACF, one needs the expression of Green's tensor of BIM plate. In general, the surface Green's tensor $\mathds G$ can be expressed in terms of Fresnel coefficients for reflection at the BIM plate \cite{JCrosse}, i.e., 
\begin{eqnarray}\label{eq10}
\mathds G(\bold r,\bold r^{\prime},\omega)&=&\frac{i}{2\pi}\int\,{d^2 k_\rho}\,\frac{e^{i\bold{k_\rho}\cdot (\bold r-\bold r^{\prime})+ik_z(z+z^{\prime})}}{k_z} r_{\mu\nu}\bold M_{\mu\nu},\nonumber\\
\end{eqnarray}
where $\bold{k_\rho}=(k_x, k_y)$ and $k_z=\sqrt{\omega^2-k_\rho^2}$ represent the wave vectors in x-y plane and z-direction, respectively; $r_{\mu\nu}=E_\mu^{ref}/E_\nu^{in}$ ($\mu,\nu \in \left\{ s, p\right\}$) stands for the reflection coefficient from $\nu$-polarized photons to $\mu$-polarized photons; the superscript  \textit{in} (\textit{ref}) simply denotes incident (reflection) photons. The explicit expressions of the matrices $\bold M_{\mu\nu}$ are given in Appendix C. In contrast to common PTS materials, the cross reflection coefficients  $r_{sp}$ and $r_{ps}$ are usually nonzero for BIMs due to the fact that magnetoelectric effect can mix s- and p- polarizations in general. 

Based on the constitutive relations and boundary conditions, one can obtain the cross-reflection coefficients \cite{ILindell} 
\begin{eqnarray}\label{rsp}
r_{sp} (r_{ps})=\frac{2\eta_0\eta c_0}{\Delta}\left[\pm i(c_+-c_-)\cos\beta- (c_++c_-)\sin\beta\right]. &{}&\nonumber\\
&{}&
\end{eqnarray}
Here, $\eta=\sqrt{\mu/\epsilon}$ ($\eta_0=\sqrt{\mu_0/\epsilon_0}$) represents the impedance of the BIM (vacuum); $c_0=\cos\theta_0$, where $\theta_0$ is the incident angle of an EM wave; $c_{\pm}=\cos\theta_{\pm}=\sqrt{k_{\pm}^2-k_\rho^2}/k_{\pm}$, where $\theta_{\pm}$ stand for refractive angles and $k_{\pm}=k(\cos\beta\pm \kappa_r)$;  $\sin\beta=\chi_r=\chi (\sqrt{\epsilon_0\mu_0}/\sqrt{\epsilon\mu})$ and $\kappa_r=\kappa (\sqrt{\epsilon_0\mu_0}/\sqrt{\epsilon\mu})$ are the relative magnetoelectric parameters; $\Delta=(\eta_0^2+\eta^2)c_0(c_++c_-)+2\eta_0\eta(c_0^2+c_+c_-)\cos\beta$. For lossless media, $k_{\pm}\geq 0$ implies the relationship $\chi_r^2+\kappa_r^2\leq 1$ \cite{RZhao}.

The key element that induces the ACF is the differential Green's functions $\Sigma_{x/z}$, which can be expressed by the cross-reflection coefficients through
\begin{eqnarray}\label{Txz}
\Sigma_{x/z}(\omega)=\frac{\omega}{2\pi}\int_{-\infty}^{\infty}d^2 k_\rho \,e^{2 i \sqrt{w^2-k_\rho^2}d}g_{x/z}(r_{sp}\mp r_{ps}).\nonumber\\
\end{eqnarray}
Here, $g_x={k_x^2}/{\sqrt{\omega^2-k_\rho^2}}$ and $g_z=\sqrt{\omega^2-k_\rho^2}$.
Substituting Eqn. \eqref{Txz} into Eqn. \eqref{eq8} and \eqref{eq9}, one can immediately obtain the ACF $F_{n}$ in an arbitrary direction $\hat n$. 

\begin{figure}[!htb]
\centering
\includegraphics[height=5.8cm, width=8.2cm, angle=0]{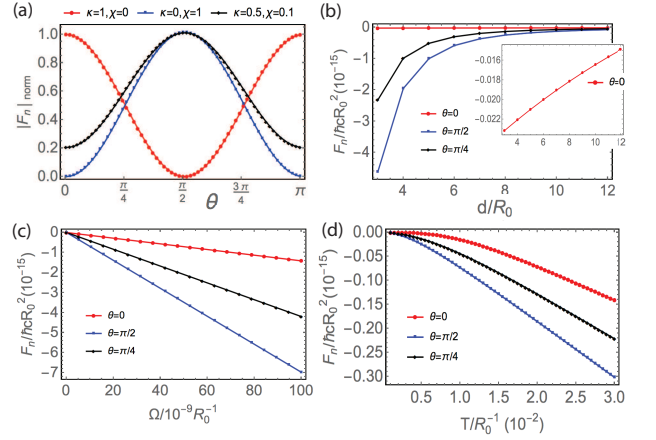}
\caption{Numerical calculation of ACF vs. angle $\theta$, rotating frequency $\Omega$ and distance $d$.
Figure (a) shows the normalized ACF $|F_n|_{norm}=|F_n/(F_n)_{max}|$ at different angles. Red, blue, and black curves correspond to Pasteur BIM ($\chi=0$, $\kappa=1$), Tellegen BIM ($\chi=1$, $\kappa=0$), and general BIM ($\chi=0.1$, $\kappa=0.5$), respectively.  Other parameters are set as: $d=12 R$, $T=0.01 R_0^{-1}$, $\Omega=10^{-9} R_0^{-1}$, $\omega_p=0.1 R_0^{-1}$, $\tau_0^{-1}=0.01 R_0^{-1}$. Figure (b) gives distance-dependence of ACF. Note that the force values are normalized by $F_0=\hbar c/R^2$. The inset of (c) gives the zoomed-in ACF at $\theta=\pi/2$. Parameters are set as:  $\kappa=-0.5$, $\chi=0.1$, and other parameters are the same as those in (a).  Figure (c) shows the frequency-dependence of ACF. Red, blue, black curves correspond to angle $\theta=0$, $\theta=\pi/2$, $\theta=\pi/4$, respectively. Parameters are set as: $d=12 R_0$, and other parameters are the same as those in (b). Figure (d) corresponds to the temperature-dependence of ACF at angles $\theta=0$, $\theta=\pi/2$, and $\theta=\pi/4$, respectively. 
\label{fig:2}}
\end{figure}

Based on the above formulas, we give the criterion for the emergence of ACF in the following:\\
(\textit{i}) For simple isotropic materials (with PS and TRS), $\chi =\kappa =0$,  and one can find $\Sigma_{x}=\Sigma_{z}=0$.  Consequently, both $F_x$ and $F_z$ vanish, leading to the vanishing of ACF in any direction. \\
(\textit{ii}) For Pasteur materials (with TRS but without PS),  $\chi =0$ and $\kappa \neq0$ lead to $\Sigma_{z}=0$ and $\Sigma_x\neq 0$. As a result, $F_x\neq 0$ and $F_z= 0$ indicate that the ACF only vanishes in the z-direction. \\
(\textit{iii}) For Tellegen materials (with PS but without TRS), $\kappa=0$ but $\chi\neq 0$, and one can show $\Sigma_x=0$ whereas $\Sigma_z\neq 0$, which results in $F_x=0$ and $F_z\neq0$. In this case, the ACF only vanishes in the x-direction. \\
(\textit{iv}) For more general cases (without PS and TRS) where $\chi\neq 0$ and $\kappa \neq 0$,  $\Sigma_x\neq 0$ and $\Sigma_z \neq 0$,  ACF can persist in any direction. 

Based on the above analysis, ACF is a general phenomenon that exists in many materials, including topological materials and chiral materials. [See Table I.]

\section{IV. Numerical Calculation and analytical limit}

We calculate the ACF numerically by considering a rotating particle described by Drude model,
where the electric permittivity is modeled by $\epsilon=\epsilon_b+\frac{\omega_p^2}{\omega(\omega+ i\tau^{-1})}$. In the formula, $\epsilon_b$ is background static electric permittivity, $\tau$ is the scattering time of electrons, and $\omega_p$ is called as plasmonic frequency.  The polarizability can be obtained from $\epsilon$, and reads as $\alpha(\omega)= 4\pi R_0^3 \epsilon_0 \frac{\epsilon-\epsilon_0}{\epsilon+2\epsilon_0}$, where $R_0$ is the radius of the particle, and $\epsilon_0$ is the vacuum permittivity. For simplicity, we set $\eta=\eta_0$ in the calculation. The numerical results are summarized in Figure 2. In Figure 2 (a), the angle-dependent ACF is shown, where L-ACF $F_x$ exists at $\kappa\neq 0$, whereas V-ACF $F_z$ exists at $\chi\neq 0$. In general case where $\chi\neq 0$ and $\kappa \neq 0$,  both $F_x$ and $F_z$ exist. Figure 2 (b) shows the distance dependence of ACF at different angles. One can see that the decaying behavior of ACF dependents on the angle $\theta$. At $\theta=0$, the L-ACF is two-orders smaller than V-ACF (inset of figure 2(b)). Figure 2 (c) gives the rotating-frequency dependence of ACF, from which one can see the ACF increases linearly with  $\Omega$. In figure 2 (d), the temperature-dependence of ACF is  shown, where ACF increase with $T$ non-linearly at low temperature, but linearly at high temperature.

We analytically obtain ACF in the low-frequency limit $\omega d \rightarrow 0$, which can be fulfilled at low temperatures due to the differential distribution function $N(T,\omega)$.   For Tellegen materials, one has $\chi=1$ and $\kappa=0$ and $r_{sp}=r_{ps}=-1$, which leads to $\Sigma_x=0$ and $\mathrm{Im}\left\{\Sigma_z\right\}=-\omega/2 d^3$.  By contrast, for Pasteur materials, $\chi=0$ and $\kappa=1$ results in $r_{sp}=-r_{ps}=-2 i\sqrt{1-s^2}/(2\sqrt{1-s^2}+\sqrt{4-s^2})$, and leads to $\Sigma_z=0$ and $\mathrm{Im}\left\{\Sigma_x\right\}=-4\omega^4/3+3\pi\omega^5d/4$. Further, if $\omega \ll \omega_p$, one can have the imaginary polarizability $\rm Im\{\alpha\}\approx - {12\pi \omega}/{(\omega_p^2\tau)}$. Under realistic conditions, when the rotating frequency is much smaller than plasmonic frequency, one can approximately obtain $\rm Im\{\alpha(\omega_+)\}N(\omega_+)-Im\{\alpha(\omega_-)\}N(\omega_-)\approx 2\Omega \,Im\{\alpha\}\partial_{\omega}N(\omega,T)+\Omega^3\,Im\{\partial_{\omega}\alpha\} \left[\partial_{\omega}^2 N(\omega,T)\right]$ assuming $T_1=T_2=T$.  By substituting these approximations into the ACF expression Eqn. \eqref{eq9}, one can readily obtain the analytical expressions of V-ACF and L-ACF:
\begin{eqnarray}
\label{acffz}F_z&=&-\frac{\hbar R_0^3}{\omega_p^2\tau d^3}\left[4\pi^2T^2\Omega+\Omega^3\right]\\
F_x&=&-\frac{16\hbar R_0^3}{\omega_p^2\tau}\left[240\zeta(5) T^5 \Omega+16\zeta(3)T^3\Omega^3\right.\nonumber
\\&&\quad\quad\quad\quad+\left.\frac{2 T}{5} \Omega^5+\frac{\mathrm{Sign}[\Omega]}{30}\Omega^6\right]
\end{eqnarray}
Note that we set parameters $\chi\rightarrow 1$ ($\chi\rightarrow 0$) and $\kappa\rightarrow 0$ ($\kappa\rightarrow 1$) in obtaining analytical result of laterally (vertically) ACF. In figure 3 we compare our analytical result with numerical results showing consistence at low temperature. From the analytical expressions, one can understand why ACF depends on rotating frequency linearly and non-linearly on temperature at low temperatures. 
\begin{figure}[!htb]
\centering
\includegraphics[height=3.3cm, width=8.4cm, angle=0]{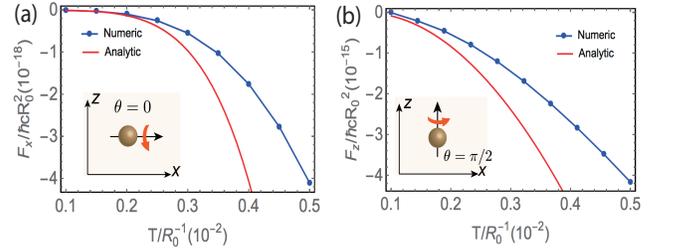}
\caption{Comparison  between Numerical calculation and analytical calculation of ACF. Figure (a) and (b) show the L-ACF and V-ACF, where analytical result (red curve) and numerical result (blue curve) are consistent at low temperature. The distance is set $d=5 R_0$, while other parameters are the same as that in Figure 2.  \label{fig:3}}
\end{figure}

\section{V. Magnetic contribution}

Our previous calculation is based on the dipolar approximation, i.e., the size of the particle is much smaller than the cut-off wave length of photons $1/\omega_p$ and the distance $d$. For a large object, the dipolar approximation is not valid, and one should use the scattering-matrix method to calculate the Casimir force \cite{ACanaguier-Durand}. Notice that, the magnetic contribution is neglected in our previous calculation. The magnetic contribution can be easily included by replacing the electric polarization $\alpha$ with the magnetic polarization $\beta$.  Also, the electric Green's tensors $\mathcal G_{ij}$ should be replaced by the magnetic Green's tensors $\mathcal H_{ij}$ which can be easily obtained from the electric counterparts by swapping polarization indices, i.e. $\mathcal H_{ij}=\mathcal G_{ij}(s\leftrightarrow p)$. 

We calculate the magnetic polarizability, and show that, for small particles, the magnetic contribution is vanishingly small.  The magnetic polarizability for a spherical particle with radius $R$ is given by
$\beta(\omega)=-R^3\left[\frac{1}{2}-\frac{3}{2(\epsilon-1)\omega^2 R^2}+\frac{3}{2(\epsilon-1)\omega R}cot( \sqrt{\epsilon-1} \omega R)\right]$  \cite{AVolokitin2}. With the magnetic polarizability and magnetic Green's tensor, one can calculate the magnetic contribution of the ACF. In figure 4, we numerically calculated the ratio between the electric contribution and the magnetic contribution. Our result shows that the magnetic contribution to the ACF is vanishingly small. In fact, the small magnetic contribution can be understood analytically. In the small particle limit ($\omega_p\,R \ll 1$), the magnetic polarizability becomes $\beta(\omega)\rightarrow -R^3(\omega_p R)^2/30$, and the magnetic and the electric polarization ratio $\beta(\omega)/\alpha(\omega) \ll 1$.   This is quite different from the large metallic sphere case where magnetic fluctuations contribute the same order as the electric counterpart \cite{milton1,milton2}.

\begin{figure}[!htb]
\centering
\includegraphics[height=3.6cm, width=5.4cm, angle=0]{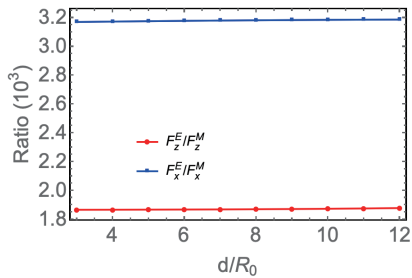}
\caption{Magnetic contribution to the ACF. Red (blue) curve represents the force ratio between the electric L-ACF (V-ACF) and magnetic L-ACF (V-ACF). \label{fig:4}}
\end{figure}

\section{VI. Discussion and comments}

(1) The ACF is parallel with rotating axis, and it does not exert torque on the rotating particle. Consequently, the ACF can not induce the heat transfer between the particle and the BIM plate. Hence, we again illustrate the dissipationless nature of ACF \cite{AManjavacas1}.

(2) Let's compare the ACF with the usual Casimir-Polder force. The usual Casimir-Polder force exists only in z-direction, and depends on $r_{ss}$ and $r_{pp}$.  In the Appendix C,  we show that the usual Casimir-Polder force can be made as small as possible and even vanished in certain cases.  Also, from equation \eqref{acffz}, one can see that V-ACF decays slower that the usual Casimir force ($\propto 1/d^5$). 

(3) In this paper, the non-equilibrium effects and spatial dispersion effects are ignored in the calculation, which could become important in some circumstances \cite{DReiche}.  The non-inertial effect is also neglected in our approach, which may be interesting for further study \cite{YZeldovich}.

(4) Quantum levitation may be possible by using V-ACF.  For example, a particle with radius $R_0=10 \,\rm{nm}$ and density $\rho_0=0.1 \,\rm{g\, cm^{-3}}$ has gravitational force $F_G \approx 4\times 10^{-21}$ N. If it rotates above a BIM plate, the V-ACF can reach $F_n\approx 10 F_G$  for parameters $\Omega = 10^{-5} R_0^{-1}\approx100$ GHz, $d=3 R_0=30$ nm, $\omega_p=0.1 R_0^{-1}\approx 1$ eV, $\tau^{-1}=200$ {meV}, $T=0.03 R_0^{-1}\approx 500$ K, $\chi=1$, and $\kappa=0$. Note that the very recent experiments have already achieved a superfast rotation of nanoparticles, making the ACF within the experimental reach \cite{RReimann}.

\section{Summary}
We have identified the first dissipationless rotation-induced force in vacuum, named axial Casimir force. The axial Casimir force emerges when a particle rotating above a plate that has either time-reversal symmetry breaking or parity-symmetry breaking. Various topological materials and chiral materials are promising candidates to observe the axial Casimir force. Due to V-ACF, quantum levitation is also possible for a particle rotating nearby a BIM plate.  Furthermore, the axial Casimir force has a slower scaling law with distance, and can dominate over the common Casimir-Polder force in certain cases.

\section{Acknowledgement}
 We are grateful to T. H. Hansson for carefully reviewing our manuscript and many helpful discussions. This work was supported by the Swedish Research Council under Contract No. 335-2014-7424.  In addition, FW's work is supported by the U.S. Department of Energy under grant Contract  No. DE-SC0012567 and by the European Research Council under grant 742104.

\hspace{2mm}

\begin{center}
\textbf{Appendix A:  Derivation of  L-ACF}
\end{center}
 
In this section, we calculate the L-ACF in detail. 
The L-ACF is contributed from two pieces including the dipole fluctuation and the field fluctuation. We calculate them separately. 

\textit{Dipole fluctuation distribution---}We can compute the first term of Eqn. \eqref{eq4}:
\begin{eqnarray}\label{fxp2}
F_{x,p}=&&\langle p_i^{fl}(t)\partial_x E_i^{ind}(\bold r_0, t)\rangle\nonumber\\
=&&\int_{\infty}^{\infty} \frac{d\omega d\omega^{\prime}}{4\pi^2}\, e^{-i(\omega+\omega^{\prime})t} \langle p_i(\omega)\partial_x E_i(\bold r_0, \omega^{\prime})\rangle\nonumber\\
=&&\int_{\infty}^{\infty} \frac{d\omega d\omega^{\prime}}{4\pi^2}\, e^{-i(\omega+\omega^{\prime})t}\nonumber\\
&&\times \langle p_i(\omega)\partial_x G_{ij}(\bold r_0, \bold r_0, \omega^{\prime})p_j(\omega^{\prime})\rangle,
\end{eqnarray}
where the derivative only acts on the first component of Green function, i.e. $\partial_x G(\bold r_0,\bold r_0,\omega)=\partial_x G(\bold r,\bold r_0,\omega)|_{\bold r=\bold r_0}$. Now, comes to the important part. Based on the FDT, one can connect the quantity $\langle p_i p_j\rangle$ with atomic polarizability $\alpha_{ij}$. Since the particle is rotating, the atomic polarizability is only well defined in the rotating frame. In the expression of Eqn. \eqref{fxp2}, $p_i$ is the electric dipoles that are defined in the lab frame. In order to express electric polarizability in terms of electric dipoles $\tilde p_{i}$ in the rotating frame, one needs the coordinate transformation \cite{AManjavacas1}
\begin{eqnarray}\label{px}
&&p_x(\omega)=\tilde p_{x}(\omega)\nonumber\\&&
p_y(\omega)=\frac{1}{2}\left[ \tilde p_{y}(\omega_+)+i\tilde p_{z}(\omega_+)+\tilde p_{y}(\omega_-)-i\tilde p_{z}(\omega_-)\right]\nonumber\\&&
p_z(\omega)=\frac{1}{2}\left[-i \tilde p_{y}(\omega_+)+\tilde p_{z}(\omega_+)+i\tilde p_{y}(\omega_-)+\tilde p_{z}(\omega_-)\right]\nonumber\\
\end{eqnarray}
where $\omega_{\pm}=\omega\pm\Omega$ is the Doppler-shifted frequency with $\Omega$ denotes the rotation frequency of the particle.  In the main text, we expressed the coordinate transformation of electric dipole by the form:  $p_i(\omega)=\Lambda_{ij}^+\tilde p_j(\omega_+)+\Lambda_{ij}^0\tilde p_j(\omega)+\Lambda_{ij}^{-} \tilde p_j(\omega_-)$. Now, one can read out $\Lambda_{ij}^{\pm}$ and $\Lambda_{ij}^0$ from above equations.  Due to the translational symmetry in x, y and z direction of the surface, the surface Green function satisfies $\partial_x G_{xx}=\partial_x G_{yy}=\partial_x G_{zz}=0$. One can find the explicit expression of surface Green's tensor in \textbf{Appendix C}, where we show that only the terms $\partial_x G_{yz}$, $\partial_x G_{zy}$, $\partial_xG_{xz}$, and $\partial_xG_{zx}$ need to be calculated. (Due to the isotropic assumption of the rotating particle ($\alpha_{xy}=\alpha_{xz}=0$), we do not need to calculate $\partial_x G_{xz}$ and $\partial_x G_{xy}$, which alway appear, respectively, with $\alpha_{xy}$ and $\alpha_{xz}$ at the same time.) The L-ACF induced from dipole fluctuation is
\begin{eqnarray}\label{fxp}
F_{x,p}
=&&\int_{-\infty}^{\infty} \frac{d\omega d\omega^{\prime}}{4\pi^2}\left\{\left[\langle p_y(\omega)\partial_x G_{yz}(\bold r_0, \bold r_0, \omega^{\prime})p_z(\omega^{\prime})\rangle\right]\right.\nonumber\\
&&\left.+\left[\langle p_z(\omega)\partial_x G_{zy}(\bold r_0, \bold r_0, \omega^{\prime})p_y(\omega^{\prime})\rangle\right]\right\}\,\delta(\omega+\omega^{\prime})\nonumber\\
=&&\int_{-\infty}^{\infty} \frac{d\omega }{4\pi^2}\, \left[\langle p_y(\omega)\partial_x G_{yz}(\bold r_0, \bold r_0, -\omega)p_z(-\omega)\rangle\right]\nonumber\\
&&+\left[\langle p_z(\omega)\partial_x G_{zy}(\bold r_0, \bold r_0, -\omega)p_y(-\omega)\rangle\right].
\end{eqnarray}
Substitute Eqn. \eqref{px} into Eqn. \eqref{fxp} , and one can obtain
\begin{eqnarray}
\langle  p_y &&(\omega)  \partial_x G_{yz}(\bold r_0, \bold r_0, -\omega)p_z(-\omega)\rangle\nonumber\\
=&&\frac{1}{4}\partial_x G_{yz}^*\langle\left[{{\tilde p_{y}(\omega_+)}}+{{i\tilde p_z(\omega_+)}}+{{\tilde p_y(\omega_-)}}-i\tilde p_z(\omega_-)\right]\nonumber\\
&&\times\left[{{-i \tilde p_y(\omega_+^{\prime})}}+\tilde p_z(\omega_+^{\prime})+{{i\tilde p_y(\omega_-^{\prime})}}+{{\tilde p_z(\omega_-^{\prime})}}\right]\rangle\nonumber\\
=&&i \pi\hbar\partial_x G_{yz}^*\left\{\mathrm{Im}\left\{{{\tilde \alpha_{yy}(\omega_+)}}+{{\tilde \alpha_{zz}(\omega_+)}}\right\}\left(n(T_1,\omega_+)+\frac{1}{2}\right)\right.\nonumber\\
&&\left.-\mathrm{Im}\left\{{{\tilde \alpha_{yy}(\omega_-)}}+\tilde \alpha_{zz}(\omega_-)\right\}\left(n(T_1,\omega_-)+\frac{1}{2}\right)\right\},
\end{eqnarray}
where $\omega_\pm=\omega\pm\Omega$, $\omega_\pm^{\prime}=-\omega\pm\Omega$, and $T_1$ is the particle temperature. 

In the same way, one can obtain
\begin{eqnarray}
\langle &&p_z(\omega)\partial_x G_{zy}(\bold r_0, \bold r_0, -\omega)p_y(-\omega)\rangle\nonumber\\
=&&-i \pi\hbar\partial_x G_{zy}^*
\left\{\mathrm{Im}\left\{\tilde \alpha_{yy}(\omega_+)+\tilde \alpha_{zz}(\omega_+)\right\}\left(n(T_1,\omega_+)+\frac{1}{2}\right)\right.\nonumber\\
&&\left.-\mathrm{Im}\left\{\tilde \alpha_{yy}(\omega_-)+\tilde \alpha_{zz}(\omega_-)\right\}\left(n(T_1,\omega_-)+\frac{1}{2}\right)\right\}.
\end{eqnarray}
Several comments in order: (i) Since the electric field $\bold E(t)$ is real, the Green function $G_{ij}(t)=\langle E_i(t)E_j(0)\rangle$ is also real, i.e., $G_{ij}^*(t)=G_{ij}(t)$. Due to the expression of Green function in $\omega$-space, $G_{ij}(t)=\int \frac{d\omega}{2\pi} e^{-i\omega t}G_{ij}(\omega)$ we can obtain $G^*(\omega)=G(-\omega)$. The same reason also suggests $\alpha^*(\omega)=\alpha(-\omega)$.
(ii) When we make the simplification $\int_{-\infty}^{\infty}\mapsto\int_{0}^{\infty}$, it's not as simple as $\int_{-\infty}^{\infty}=2\int_{0}^{\infty}$. In fact, we should use the equality 
\begin{eqnarray}
\int_{-\infty}^0 &&d\omega ~\partial_x G_{yz}^{*}(\omega){\rm Im}\{\alpha(\omega_+)\}\left(n(\omega_+)+\frac{1}{2}\right)\nonumber\\
=&&\int_{0}^{\infty} d\omega\partial_x G_{yz}^{*}(-\omega){\rm Im}\{\alpha(-\omega_-)\}\left(n(-\omega_-)+\frac{1}{2}\right)\nonumber\\
=&&\int_{0}^{\infty} d\omega\partial_x G_{yz}(\omega){\rm Im}\{\alpha(\omega_-)\}\left(n(\omega_-)+\frac{1}{2}\right).
\end{eqnarray}
Note that the FDT in the rotating frame leads to the relation $\langle \tilde p_i(\omega)\tilde p_j(\omega^{\prime})\rangle=2\pi\hbar \delta(\omega+\omega^{\prime})\mathrm{Im}\left\{\tilde\alpha_{ij}(\omega)\right\}coth(\frac{\beta \omega}{2})$. 
Therefore, the Casimir force $F_{x,p}$ due to electric dipole fluctuation is
\begin{eqnarray}
F_{x,p}=&&\frac{i\hbar}{4\pi}\int_{-\infty}^{\infty} d\omega (\partial_x G_{yz}^*-\partial_x G_{zy}^*)\times\nonumber\\
&&\left\{\mathrm{Im}\left\{\tilde\alpha_{yy}(\omega_+)+\tilde\alpha_{zz}(\omega_+)\right\}\left(n(T_1,\omega_+)+\frac{1}{2}\right)\right.\nonumber\\
&&\left.-\mathrm{Im}\left\{\tilde\alpha_{yy}(\omega_-)+\tilde\alpha_{zz}(\omega_-)\right\}\left(n(T_1,\omega_-)+\frac{1}{2}\right)\right\}\nonumber
\end{eqnarray}
One can express the above result only in the frequency region $\omega \geq 0$ by using ${\rm Im}\{\tilde\alpha(-\omega)\}=\rm{Im}\{\tilde\alpha^*(\omega)\}=-{\rm Im}\{\tilde\alpha(\omega)\}$ and $\left(n(T_1,-\omega)+\frac{1}{2}\right)=-\left(n(T_1,\omega)+\frac{1}{2}\right)$:
\begin{eqnarray}
F_{x,p}
=&&\frac{\hbar}{2\pi}\int_{0}^{\infty} d\omega \mathrm{Im}\left\{\partial_x G_{yz}-\partial_x G_{zy}\right\}\times\nonumber\\
&&\left\{\mathrm{Im}\left\{\tilde\alpha_{yy}(\omega_+)+\tilde\alpha_{zz}(\omega_+)\right\}\left(n(T_1,\omega_+)+\frac{1}{2}\right)\right.\nonumber\\
&&\left.-\mathrm{Im}\left\{\tilde\alpha_{yy}(\omega_-)+\tilde\alpha_{zz}(\omega_-)\right\}\left(n(T_1,\omega_-)+\frac{1}{2}\right)\right\}\nonumber
\end{eqnarray}

\textit{Field fluctuation distribution---}One can compute the second term of Eqn. \eqref{eq4} induced by the electric field fluctuation.
\begin{eqnarray}\label{fxe}
F_{x,E}
=&&\langle {\alpha}_{ij} E_j^{fl}(\bold r_0, t) \partial_x E_i^{fl}(\bold r_0, t) \rangle\nonumber\\
=&&\int_{-\infty}^{\infty}\frac{d\omega d\omega^{\prime}}{(2\pi)^2} e^{-i(\omega+\omega^{\prime})t}\langle {\alpha}_{ij}(\omega)E_j^{fl}(\omega)\partial_x E_i^{fl}(\omega^{\prime})\rangle,\nonumber\\
\end{eqnarray}
where ${\alpha}_{ij}$ denotes the effective polarizability seen in the lab frame, corresponding to the polarizability $\tilde \alpha$ in its rotating frame via \cite{AManjavacas1}:
\begin{eqnarray}
{\alpha}_{xx}(\omega)=&& \tilde\alpha_{xx}(\omega)\nonumber\\
{\alpha}_{yy}(\omega)=&&\frac{1}{4}\left(\tilde\alpha_{yy}(\omega_+)+\tilde\alpha_{zz}(\omega_+)+\tilde\alpha_{yy}(\omega_-)+\tilde\alpha_{zz}(\omega_-)\right)\nonumber\\
=&&{\alpha}_{zz}(\omega)\nonumber\\
{\alpha}_{yz}(\omega)=&&\frac{i}{4}\left(\tilde\alpha_{yy}(\omega_+)+\tilde\alpha_{zz}(\omega_+)-\tilde\alpha_{yy}(\omega_-)-\tilde\alpha_{zz}(\omega_-)\right),\nonumber\\
=&&-{\alpha}_{zy}(\omega)
\end{eqnarray}
from which one can read out $\Gamma_{ijkl}^{\pm}$ and $\Gamma_{ijkl}^{0}$ in the main text. Substituting $\tilde{\alpha}_{yz}(\omega)$ and $\langle E_z^{fl}(\omega)\partial _x E_y^{fl}(\omega^{\prime})\rangle=\langle \partial _x E_y^{fl}(\omega^{\prime}, \bold r_0)E_z^{fl}(\omega,\bold r_0)\rangle=4\pi\delta(\omega+\omega^{\prime})\mathrm{Im}\{\partial_xG_{yz}(\bold r_0, \bold r_0,\omega^{\prime})\} \left(n(T_2,\omega^{\prime})+\frac{1}{2}\right)$ ( $T_2$ is the temperature of the surface.) into Eqn. \eqref{fxe}, one can get

\begin{eqnarray}
F_{x,E}=&&\langle p_i^{ind}(t)\partial_x E_i^{fl}(\bold r_0, t)\rangle\nonumber\\
=&&-\frac{\hbar}{2\pi}\int_{0}^{\infty} d\omega ~\mathrm{Im}\{\partial_x G_{yz}-\partial_xG_{zy}\}\nonumber\\
&&\times\mathrm{Im}\{\tilde\alpha_{yy}(\omega_+)+\tilde\alpha_{zz}(\omega_+)-\tilde\alpha_{yy}(\omega_-)-\tilde\alpha_{zz}(\omega_-)\}\nonumber\\
&&\times\left[n(T_2,\omega)+\frac{1}{2}\right].
\end{eqnarray}
Combine the force induced by dipole fluctuation and field fluctuation distribution, and we can obtain the final expression of ACF:
\begin{eqnarray}
F_x=&&\frac{\hbar}{2\pi}\int_{0}^{\infty} d\omega\, \mathrm{Im}\left\{\partial_x G_{yz}-\partial_x G_{zy}\right\}\nonumber\\
&&\left[\mathrm{Im}\left\{\tilde\alpha_{yy}(\omega_+)+\tilde\alpha_{zz}(\omega_+)\right\}N(\omega_+)\right.\nonumber\\
&&\left.- \mathrm{Im}\left\{\tilde\alpha_{yy}(\omega_-)+\tilde\alpha_{zz}(\omega_-)\right\}N(\omega_-)\right],
\end{eqnarray}
where $N(\omega_\pm)=n(T_1, \omega_\pm)-n(T_2, \omega)$. Even at zero temperature $T_1=T_2=0$, $N(\omega_\pm)\neq0$ which indicates that the ACF is totally contributed from quantum fluctuation. By assuming the particle is isotropic, i.e., $\alpha_{ij}=\alpha \delta_{ij}$, the formula in the main text is obtained.

\vspace{2mm}

\begin{center}
\textbf{Appendix B: Derivation of Axial Casimir force (ACF) in Arbitrary Direction.}
\end{center}

In this section, we show how to calculate the  ACF of a particle rotating along an arbitrary axis $\hat n=(\cos\theta, 0,\sin\theta)$. We calculate the Casimir force in x, y, z directions, respectively, and project them along the rotating axis.  $(p_x^{\theta}, p_y^{\theta}, p_z^{\theta})$ represents electric dipole of the particle in the lab frame, which can be obtained via coordinate rotation:
\begin{eqnarray}
\left(\begin{array}{cc}
p_x^{\theta}\\ p_z^{\theta}
\end{array}\right)=
\left(\begin{array}{cc}
\cos\theta & -\sin\theta\\
\sin\theta & \cos\theta
\end{array}\right)\left(\begin{array}{cc}
p_x\\ p_z
\end{array}\right),
\end{eqnarray}
where $p_x$ and $p_z$ are the electric dipoles that are obtained in last section. The polarization at $\hat n=(\cos\theta, 0, \sin\theta)$ corresponds to the polarization at $\hat n=(1,0,0)$ via 
\begin{eqnarray}
\alpha_{xx}^{\theta}=&&\cos^2\theta {\alpha}_{xx}+\sin^2\theta {\alpha}_{zz}; \,
\alpha_{yy}^{\theta}=\langle p_y p_y\rangle= {\alpha}_{yy};  \nonumber\\
\alpha_{zz}^{\theta}=&&\cos^2\theta  {\alpha}_{zz} +\sin^2\theta \alpha_{xx} ;\,
\alpha_{xy}^{\theta}=-\sin\theta  {\alpha}_{zy};   \nonumber\\
\alpha_{yx}^{\theta}=&&-\sin\theta {\alpha}_{yz}; 
\alpha_{xz}^{\theta}=\sin\theta \cos\theta  {\alpha}_{xx}-\sin\theta \cos\theta  {\alpha}_{zz} \nonumber\\
\alpha_{zx}^{\theta}=&&\alpha_{xz}^{\theta}; \,
\alpha_{yz}^{\theta}=\cos\theta  {\alpha}_{yz}; \,
\alpha_{zy}^{\theta}=\cos\theta {\alpha}_{zy},
\end{eqnarray}
In the following, we obtain the Casimir force (induced by rotation) in the x-direction $F_x$ and in the z-direction $F_z$, respectively. Then, the total ACF is \begin{eqnarray}
F_n(\theta)=F_x(\theta)\cos\theta+F_z(\theta)\sin\theta.
\end{eqnarray} 

\textit{The axial Casimir force in the x-direction---} Notice that the L-ACF calculated here is different from that in \textbf{Appendix A}. Because the rotating axis is not parallel with the plate anymore, i.e., $F_x(\theta)\neq F_x$, thus one needs to re-calculate the L-ACF in this case. Again, the L-ACF is induced from two parts contributions: (i) the electric dipole fluctuation and (ii) the electric field fluctuation, i.e.,
$F_x(\theta)=F_{x,p}+F_{x,E},$
where
\begin{eqnarray}
F_{x,p}=&&\int_{-\infty}^{\infty}\frac{d\omega d\omega^{\prime}}{(2\pi)^2}e^{-i(\omega+\omega^{\prime})t}\nonumber\\
&&\times\langle p_i^{\theta}(\omega)\partial_x G_{ij}(\bold r_0,\bold r_0, \omega^{\prime}) p_j^{\theta}(\omega^{\prime})\rangle;\\
F_{x,E}=&&\int_{-\infty}^{\infty}\frac{d\omega d\omega^{\prime}}{(2\pi)^2} e^{-i(\omega+\omega^{\prime})t}\langle  \alpha^{\theta}_{ij}(\omega)E_j^{fl}(\omega)\partial_x E_i^{fl}(\omega^{\prime})\rangle.\nonumber\\
\end{eqnarray}
Substituting the electric dipole into the expression of the dipole-induced Casimir force, and one can obtain
\begin{eqnarray}
F_{x,p}=\frac{\hbar}{2\pi}\int_0^{\infty}d\omega \left[\sin\theta \cos\theta f_{p1}+\cos\theta f_{p2}-\sin\theta f_{p3}\right],\nonumber\\
\end{eqnarray}
where 
\begin{eqnarray}\label{fp1}
f_{p1}=&&{\rm Re}\,\left\{(\partial_xG_{xz}+\partial_xG_{zx})\right\}\nonumber\\
&&\times\left\{4\,{\rm Im}\{\tilde \alpha_{xx}(\omega) \}\left(n(T_1,\omega)+\frac{1}{2}\right)\right.\nonumber\\
&&-{\rm Im}\left\{\tilde \alpha_{yy}(\omega_{+})+\tilde \alpha_{zz}(\omega_+)\right\}\left(n(T_1,\omega_+)+\frac{1}{2}\right)\nonumber\\
&&\left.-{\rm Im}\left\{\tilde \alpha_{yy}(\omega_{-})+\tilde \alpha_{zz}(\omega_-)\right\}\left(n(T_1,\omega_-)+\frac{1}{2}\right)\right\}.\nonumber\\
\end{eqnarray}
\begin{eqnarray}\label{fp2}
f_{p2}=&&{\rm Im}\left\{(\partial_xG_{yz}-\partial_xG_{zy})\right\}\nonumber\\
&&\times\left\{{\rm Im}\left\{\tilde \alpha_{yy}(\omega_{+})+\tilde \alpha_{zz}(\omega_+)\right\}\left(n(T_1,\omega_+)+\frac{1}{2}\right)\right.\nonumber\\
&&\left.-{\rm Im}\left\{\tilde \alpha_{yy}(\omega_{-})+\tilde \alpha_{zz}(\omega_-)\right\}\left(n(T_1,\omega_-)+\frac{1}{2}\right)\right\};\nonumber\\
\end{eqnarray}
\begin{eqnarray}
\label{fp3}
f_{p3}=&&{\rm Im}\left\{(\partial_xG_{yx}-\partial_xG_{xy})\right\}\nonumber\\
&&\times\left\{{\rm Im}\left\{\tilde \alpha_{yy}(\omega_{+})+\tilde \alpha_{zz}(\omega_+)\right\}\left(n(T_1,\omega_+)+\frac{1}{2}\right)\right.\nonumber\\
&&\left.-{\rm Im}\left\{\tilde \alpha_{yy}(\omega_{-})+\tilde \alpha_{zz}(\omega_-)\right\}\left(n(T_1,\omega_-)+\frac{1}{2}\right)\right\}.\nonumber\\
\end{eqnarray}
The electric fluctuating field contribution to the lateral Casimir force is
\begin{eqnarray}\label{fye}
F_{x,E}=\frac{\hbar}{2\pi}\int_0^{\infty} d\omega \left[\sin\theta \cos\theta f_{E1}+\cos\theta f_{E2}- \sin\theta f_{E3}\right],\nonumber\\
\end{eqnarray}
where 
\begin{eqnarray}\label{fE1}
f_{E1}
=&& \rm {Im}\left\{\partial_x G_{xz}+\partial_xG_{zx}\right\} {\rm Re}\left\{ 4\tilde \alpha_{xx}\right.-\left(\tilde \alpha_{yy}(\omega_+)
+\tilde \alpha_{zz}(\omega_+)\right.\nonumber\\
&&\left.\left.+\tilde \alpha_{yy}(\omega_-)+\tilde \alpha_{zz}(\omega_-)\right)\right\}\times\left(n(T_2,\omega)+\frac{1}{2}\right)\nonumber\\
f_{E2}=&&-{\rm Im}\{\partial_x G_{yz}-\partial_x G_{zy}\}\,{\rm Im}\left\{\tilde \alpha_{yy}(\omega_+)+\tilde \alpha_{zz}(\omega_+)\right.\nonumber\\
&&\left.-\tilde \alpha_{yy}(\omega_-)-\tilde \alpha_{zz}(\omega_-)\right\}
\times\left(n(T_2,\omega)+\frac{1}{2}\right)\nonumber\\
f_{E3}
=&&-{\rm Im}\{\partial_x G_{yx}-\partial_x G_{xy}\}\,{\rm Im}\left\{\tilde \alpha_{yy}(\omega_+)+\tilde \alpha_{zz}(\omega_+)\right.\nonumber\\
&&\left.-\tilde \alpha_{yy}(\omega_-)-\tilde \alpha_{zz}(\omega_-)\right\}\left(n(T_2,\omega)+\frac{1}{2}\right)
\end{eqnarray}
Therefore, the total L-ACF in x direction is
\begin{eqnarray}
F_x(\theta)=&&\frac{\hbar}{2\pi}\int_0^{\infty} d\omega \left[\sin\theta \cos\theta \left(f_{p1}+ f_{E1}\right)\right.\nonumber\\
&&\left.+\cos\theta \left(f_{p2}+f_{E2}\right)- \sin\theta \left(f_{p3}+f_{E3}\right)\right]
\end{eqnarray}

\textit{The Casimir force in the y-direction $F_{y}$---} The force in the y-direction has the same form as that in the x-direction. The only difference is that all derivatives on Green tensors changes from $\partial_x G_{ij}$ to $\partial_y G_{ij}$.  If we let $\theta=0$, the expression coincides with the expressions in reference \cite{AManjavacas}.

\textit{The Casimir force in the z-direction $F_z$---} The form of $F_z$  is different from $F_x$ and $F_y$ due to the non-vanishing diagonal terms $\partial_zG_{ii}$.  We can write the Casimir force in the form
$F_{z}^{tot}(\theta)=F_z^d(\theta)+F_z(\theta)$,
where $F_z^d(\theta)$ is the Casimir force due to the diagonal terms $\partial_zG_{ii}$, and $F_z(\theta)$ is the rotation-induced Casimir force. $F_z(\theta)$ has the similar form as $F_x(\theta)$ and $F_y(\theta)$, and one can obtain $F_z(\theta)$ by the substitution  $\partial_{x/y} G_{ij}\rightarrow \partial_z G_{ij}$. The diagonal Casimir force $F_z^d(\theta)$ corresponds to the usually referred Casimir-Polder force, whereas the off-diagonal Casimir force is induced by rotation. 

The total ACF along the rotating axis is 
\begin{eqnarray}\label{appfn}
F_n=&&F_x(\theta) \cos\theta+F_z(\theta) \sin\theta\nonumber\\
\approx&& F_x\cos^2\theta +F_z\sin^2\theta,
\end{eqnarray}
where $F_x\equiv F_x(\theta=0)$ and $F_z\equiv F_z(\theta=\pi/2)$.  In deriving Eqn. \eqref{appfn}, we have used the approximation that $\Omega/\omega \ll 1$.  We also use the fact that $f_{p2}+f_{E2}$ and $f_{p3}+f_{E3}$ vanish for $F_z(\theta)$ and $F_x(\theta)$, respectively. (The reason relies on the Green's tensor form in Appendix C.) Note that we derived the formula Eqn. \eqref{eq8}  announced in the main text. 
\\

In the following, one can derive the diagonal term $F_z^d(\theta)$ by considering the dipole fluctuation and the field fluctuation, respectively. 

The Casimir force in the z-direction induced by dipole fluctuation is 
\begin{eqnarray}
F_{z,p}^d=&&\int_{-\infty}^{\infty}\frac{d\omega d\omega^{\prime}}{(2\pi)^2}e^{-i(\omega+\omega^{\prime})t}\left\{\langle p_x^{\theta}(\omega)\partial_z G_{xx}(\bold r_0,\bold r_0,\omega^{\prime}) p_x^{\theta}(\omega^{\prime})\rangle\right.\nonumber\\
&&\left.+\langle p_y^{\theta}\partial_z G_{yy} p_y^{\theta}\rangle+\langle p_z^{\theta}\partial_z G_{zz} p_z^{\theta}\rangle\right\}
\end{eqnarray}

The Casimir force in the z-direction induced by field fluctuation is
\begin{eqnarray}
F^d_{z,E}
=&&\int_{-\infty}^{\infty}\frac{d\omega d\omega^{\prime}}{(2\pi)^2} e^{-i(\omega+\omega^{\prime})t} (4\pi \hbar\delta(\omega+\omega^{\prime}))\left\{\alpha_{xx}^{\theta} {\rm Im}\{ \partial_z G_{xx}\}\right.\nonumber\\
&&\left.+ \alpha_{yy}^{\theta} {\rm Im}\{\partial_z G_{yy}\}+\alpha_{zz}^{\theta} {\rm Im}\{ \partial_z G_{zz}\}\right\}\left(n(T_2,\omega)+\frac{1}{2}\right).\nonumber\\
\end{eqnarray}
Unlike the the case rotation-induced Casimir force, the signs before the integral in the expressions of $F^d_{z,p}$ and $F^d_{z,E}$ are the same. This is definitely reasonable, meaning that, even without rotation, these two terms still exist. 
Add up the dipole contribution and the field contribution, We get the total diagonal Casimir force in the z-direction:
\begin{eqnarray}
F_z^d=&&F_{z,p}^d+F_{z,E}^d\nonumber\\
\approx &&\frac{2\hbar}{\pi}\int_0^{\infty}d\omega\, \mathrm{Im}\left\{\partial_z(G_{xx}+G_{yy}+G_{zz})\times\alpha(\omega)\right\}\nonumber\\
&&\times \left[n(T,\omega)+\frac{1}{2}\right],
\end{eqnarray}
where following assumptions is implied: equal temperature $T_1=T_2$, isotropic polarizability $\alpha_{ij}=\alpha\,\delta_{ij}$, and $\left[\alpha(\omega_+)\left(n(T,\omega_+)+\frac{1}{2}\right)+\alpha(\omega_-)\left(n(T,\omega_-)+\frac{1}{2}\right)\right]\approx 2\alpha(\omega)\left[n(T,\omega)+\frac{1}{2}\right]$ for $\Omega \ll k_B T$.  In the limit of $T\rightarrow 0$, the diagonal Casimir force $F_z^d=\frac{\hbar}{\pi} \int_0^{\infty} d\omega\,\, \mathrm{Tr}\left[\alpha_{ij}\partial_zG_{ij}\right]$, which agrees with the Casimir-Polder formula in Ref. \cite{FIntravaia}. According to \textbf{Appendix C}, the differential Green's function reads  $\partial_z(G_{xx}+G_{yy}+G_{zz})=r_{ss}k^2+r_{pp}(k_\rho^2-k_z^2)$. For a metallic surface, $r_{ss}=-r_{pp}$, and $ \partial_z(G_{xx}+G_{yy}+G_{zz})=-\frac{1}{2\pi}\int dk_x dk_y\,e^{2i k_z z}\,r_{ss}(2k_z^2)$, consequently, the above formula become consistent with the Casimir-Polder formula in Refs. \cite{AVolokitin2} and  \cite{FIntravaia}. 
%

\begin{center}
\textbf{Appendix C: Surface Green's Tensor For BIM Plate.}
\end{center}

\textit{ General expression of surface Green's tensor---}The surface Green's tensor can be derived from the knowledge of Fresnel coefficients, i.e., $r_{ss}$ and $r_{pp}$ in usual cases, where $r_{ss}$ ($r_{pp}$) stands for the reflection coefficients from TE(TM) wave to TE(TM) wave \cite{lnovothy}. However, when there is a mix between TE wave and TM wave, the surface Green's tensor also depends on $r_{sp}$ and $r_{ps}$ \cite{JCrosse}, where $r_{ps}$ ($r_{sp}$) stands for the cross-reflection coefficients from TE (TM) wave to TM (TE) wave. We then give the general expression of the surface Green function by taking $r_{ps}$ ($r_{sp}$) into consideration
\begin{eqnarray}
\mathds G(\bold r,\bold r^{\prime},\omega)&&=\frac{i}{2\pi}\int\,\frac{d k_x dk_y}{k_z}\,e^{ik_x(x-x^{\prime})}e^{ik_y(y-y^{\prime})}e^{ik_z(z+z^{\prime})}\nonumber\\
&&\times\left[r_{ss}M_{ss}+r_{pp}M_{pp}+r_{sp}M_{sp}+r_{ps}M_{ps}\right],
\end{eqnarray}
where \cite{JCrosse}
\begin{eqnarray}
{M}_{ss}&=&
\left(\begin{array}{ccc}
\frac{k_y^2}{k_\rho^2}k^2&-\frac{k_xk_y}{k_\rho^2}k^2 & 0\\
-\frac{k_xk_y}{k_\rho^2}k^2 & \frac{k_x^2}{k_\rho^2}k^2 & 0\\
0 & 0 & 0
\end{array}\right),
\\
{M}_{pp}&=&
\left(\begin{array}{ccc}
-\frac{k_x^2k_z^2}{k_\rho^2}&-\frac{k_xk_yk_z^2}{k_\rho^2} & -k_xk_z\\
-\frac{k_xk_yk_z^2}{k_\rho^2} & -\frac{k_y^2k_z^2}{k_\rho^2} & -k_yk_z\\
k_xk_z & k_yk_z & k_\rho^2
\end{array}\right),
\\
{M}_{sp}&=&
\left(\begin{array}{ccc}
\frac{k_xk_yk_z}{k_\rho^2}k&\frac{k_y^2k_z}{k_\rho^2}k & k_yk\\
-\frac{k_x^2k_z}{k_\rho^2}k & -\frac{k_xk_yk_z}{k_\rho^2}k & -k_xk\\
0 & 0 & 0
\end{array}\right),
\\
{M}_{ps}&=&
\left(\begin{array}{ccc}
-\frac{k_xk_yk_z}{k_\rho^2}k&\frac{k_x^2k_z}{k_\rho^2}k & 0 \\
-\frac{k_y^2k_z}{k_\rho^2}k & \frac{k_xk_yk_z}{k_\rho^2}k & 0 \\
k_y k & -k_xk & 0
\end{array}\right).
\end{eqnarray}
And the Fresnel reflection coefficients are given by \cite{ILindell}
\begin{eqnarray}
r_{ss}=&&\frac{1}{\Delta}\left\{(\eta^2-\eta_0^2)c_0(c_++c_-)+2\eta_0\eta (c_0^2-c_{+}c_{-})\cos\beta\right\};\nonumber\\
r_{pp}=&&\frac{-1}{\Delta}\left\{(\eta^2-\eta_0^2)c_0(c_++c_-)-2\eta_0\eta (c_0^2-c_{+}c_{-})\cos\beta\right\};\nonumber\\
r_{sp}=&&\frac{2\eta_0\eta c_0}{\Delta}\left[i(c_{+}-c_{-})\cos\beta-(c_{+}+c_{-})\sin\beta\right];\nonumber\\
r_{ps}=&&\frac{-2\eta_0\eta c_0}{\Delta}\left[i(c_{+}-c_{-})\cos\beta+(c_{+}+c_{-})\sin\beta\right],
\end{eqnarray}
Corresponding definitions e.g., $k_\rho$, $\Delta$, etc. are the same as those in the main text. 
 Let's check the trivial case for an ideal metal plate,  where $\chi=\kappa=0$, and $\eta=\sqrt{\mu/\epsilon}\rightarrow 0$. In this case, $r_{pp}=-r_{ss}=1$ and $r_{sp}=r_{ps}=0$. 
In another interesting case, by assuming $\eta \mapsto \eta_0$ and $\beta \mapsto \pi/2$ (perfect Tellegen Materials), $r_{ss}=r_{pp}=0$ leads to the vanishing of the usual Casimir-Polder Force.

\begin{figure*}[!htb]
\centering
\includegraphics[height=4.8cm, width=10.6cm, angle=0]{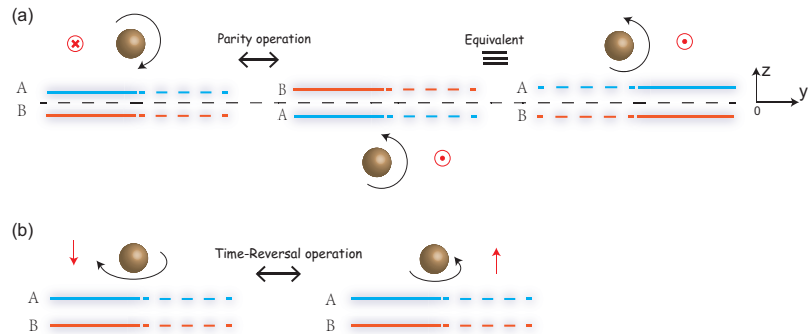}
\caption{Symmetry analysis of L-ACF and V-ACF. Figure (a) and (b) show that the L-ACF and V-ACF must, respectively vanish for materials with PS and TRS. $\odot$ and $\otimes$  represent the rotating direction of the particle, perpendicular out of plane or into plane. $\uparrow$ and $\downarrow$ mean that the rotating direction of the particle is upward or downward. \label{figuresupp1}}
\end{figure*}

\textit{ Symmetry Analysis---}We analyze the TRS and PS of the system, and demonstrate why the L-ACF (V-ACF) vanishes for systems with PS (TRS).
Our demonstration is based on Figure \ref{figuresupp1} (a) and (b). We use four kinds of style lines (solid and dashed, blue and red) to represent the plate, in order to capture the chiral nature of the plate. With our definition of ACF (Eqn. (3) and (4) in the main text), it must satisfy 
\begin{eqnarray}
F_{n}(\Omega)=-F_n(-\Omega).
\end{eqnarray}
When the rotating axis is parallel with the plate, the parity operation (with respect to $z=0$ plane) will transform left figure of (a) to the middle figure of (a), which is equivalent to right figure of (a). One can find that, if there is parity symmetry of the plate (solid lines = dashed lines), then one has $F_x(\Omega)=F_x(-\Omega)$.  Therefore, the only solution of this equation is $F_x(\Omega)=0$, i.e., L-ACF vanishes.  Analogously, if there is time-reversal symmetry of the plate, one can show that $F_z(\Omega)=F_z(-\Omega)$. And the only solution is $F_z(\Omega)=0$, i.e., V-ACF vanishes.

\vspace{2mm}

\begin{center}
\textbf{Appendix D: Derivation of analytical limit of axial Casimir force}
\end{center}

In this part, we give the detail derivation of the ACF for Tellegen and Pasteur materials in the low frequency limit $\omega d\rightarrow 0$. 

For Tellegen materials, $\chi\rightarrow 1$ and $\kappa\rightarrow 0$. Thus, the reflection coefficients can be obtained 
$r_{sp}=r_{ps}\rightarrow -1$ leading to $\Sigma_x=0$ and
\begin{eqnarray}
\Sigma_z
=-2\omega^4 \int_0^{\infty} ds (s\sqrt{1-s^2}\,\,e^{2 i \sqrt{1-s^2}\omega d})
\end{eqnarray}
In the limit $\omega d\rightarrow 0$, one can obtain $\mathrm{Im}\left\{\Sigma_z\right\}\approx-{\omega}/{2d^3}$.

For Pasteur materials, $\chi\rightarrow 0$ and $\kappa\rightarrow 1$. Thus, the reflection coefficients can be obtained 
$r_{sp}=-r_{ps}\rightarrow \frac{- 2 i \sqrt{1-s^2}}{2\sqrt{1-s^2}+\sqrt{4-s^2}}$ leading to $\Sigma_z=0$ and
\begin{eqnarray}
\Sigma_x
=\int_{0}^{\infty} ds\,\,\frac{s^3~e^{2 i \sqrt{1-s^2}\omega d}}{\sqrt{1-s^2}}\left[\frac{{(-2 i)\omega^4}~\sqrt{1-s^2}}{2\sqrt{1-s^2}+\sqrt{4-s^2}}\right].
\end{eqnarray}
In the limit $\omega d\rightarrow 0$, one can obtain $\mathrm{Im}\left\{\Sigma_x\right\}\approx-4 \omega^4/3+3\pi \omega^5 d/4\approx -4 \omega^4/3$.
Under the assumption $\omega \ll \omega_p$, the imaginary part of the polarizability reads $\mathrm{Im} \left\{\alpha(\omega)\right\}\approx -12\pi R_0^3 \omega/\omega_p^2\tau$.

\end{document}